\documentclass[twocolumn,prl,showpacs,preprintnumbers,aps,10pt]{revtex4-2}
\usepackage{graphicx,latexsym,amssymb,amsmath,color,multirow,mathrsfs,booktabs,ifpdf,epsfig,dcolumn,bm,longtable}
\usepackage{changes}
\usepackage{hyperref}
\hypersetup{colorlinks=true,linkcolor=blue,filecolor=magenta,urlcolor=cyan}

\begin{document}
	
\title{Neutron skin impurity from Coulomb core polarization in $^{208}\text{Pb}$:\\ Insights from PREX-II and validation via the $(^{3}\text{He}, t)$ isobaric analog state reaction}

\author{Phan Nhut Huan} 
\email{phannhuthuan@duytan.edu.vn}
\affiliation{Institute of Fundamental and Applied Sciences, Duy Tan University, Ho Chi Minh City 70000, Vietnam}
\affiliation{Faculty of Natural Sciences, Duy Tan University, Da Nang City 50000, Vietnam}

\begin{abstract}
I investigate the effect of Coulomb core-polarization in $^{208}\text{Pb}$, highlighting its impact on uncertainties in neutron distribution measurements due to the discrepancy between neutron excess and neutron skin regions at the surface. This effect is validated using the $(^{3}\text{He}, t)$ isobaric analog state (IAS) reaction at 420 MeV. My findings reveal that the Coulomb boundary radius, where core-polarization becomes negligible, is critical for accurately probing the neutron skin region in the target nucleus. Consequently, I propose extending the $^{208}\text{Pb}(^{3}\text{He}, t)$IAS reaction experiment conducted by Zegers et al. [Phys. Rev. Lett. 99, 202501 (2007)] to measure cross sections up to 3$^\circ$ angles to more precisely determine the neutron skin thickness in $^{208}\text{Pb}$ from theoretical reaction analysis. Additionally, I observe that, under the assumption of a large neutron skin in $^{208}\text{Pb}$, as suggested by the PREX-II result, the core-polarization effect introduces impurity into the neutron skin region. \\
\end{abstract}


\maketitle

Comprehending nucleon distribution within atomic nuclei is essential for advancing nuclear physics, exploring the equation of state (EOS) of neutron-rich matter, and addressing astrophysical phenomena \cite{Nunes20}. Accurate proton and neutron distributions are vital, with neutron skin thickness in heavy nuclei offering insights into neutron star properties \cite{Brown00}. Probing these distributions is challenging due to differing techniques for protons and neutrons. Protons are measured using electron scattering, providing charge density distributions crucial for nuclear structure studies. In contrast, neutron distributions are evaluated through proton-nucleus elastic scattering, as exhibited in the RCNP experiment, which measured the neutron skin thickness in $^{208}\text{Pb}$ as $0.211^{+0.054}_{-0.063}$ fm \cite{Zenihiro10}. Nevertheless, parity-violating electron scattering (PVES) in the Lead Radius Experiment (PREX) is the most promising model-independent technique. PREX-I \cite{PREX} and PREX-II \cite{PREX-II} at Jefferson Laboratory measure neutron distribution in $^{208}\text{Pb}$ using PVES, leveraging the weak force's stronger coupling to neutrons. PREX-II result reports a neutron skin thickness of $0.283 \pm 0.071$ fm. Furthermore, reanalysis of PREX-II using contemporary relativistic and non-relativistic energy density functionals (EDFs) predicts a neutron skin thickness of $0.19 \pm 0.02$ fm, and a symmetry energy slope of $54 \pm 8$ MeV, aligning with astrophysical observations and providing stringent constraints on nuclear models \cite{Reinhard21}.

This study addresses significant uncertainties in $^{208}\text{Pb}$ neutron distribution measurements. A major contributor to these uncertainties is the often-overlooked effect of Coulomb core polarization. For nuclei in the ground state with $N > Z$, where the $Z$ neutrons and $Z$ protons are considered as the core, Auerbach and Van Giai \cite{Auerbach81} investigated the connection between the density distributions described by $\rho_n(r) - \rho_p(r) = \rho_{nexc}(r) + \delta \rho(r)$, where $\rho_n(r) - \rho_p(r)$ represents the difference between neutron and proton densities (isovector density), directly related to the neutron skin thickness $\Delta R_{np} = R_n -R_p$, which $R_{n(p)}$ is the root mean square (rms) radius of the neutron (proton) distribution. The term $\rho_{nexc}(r) = \sum_{i = Z+1}^{N} \left|\varphi_i^n (r)\right|^2$ denotes the density of the $N-Z$ excess neutrons, while $\delta \rho(r) = \sum_{i = 1}^{Z} \left|\varphi_i^n (r)\right|^2 - \sum_{i = 1}^{Z} \left|\varphi_i^p (r)\right|^2$ reflects the density difference between the $Z$ neutrons and $Z$ protons in the core, accounting for the core-polarization effect. In this context, $\varphi_i^{n(p)}(r)$ represents the single-particle wave function of the neutron (proton). The Coulomb force induces core polarization by pushing protons outward, resulting in a $\delta \rho(r)$ with a nodal structure characterized by positive values inside the nucleus and negative values outside. This effect is pronounced when the amount of neutron excess is small, particularly in nuclei with a significant Coulomb potential. The neutron excess thickness (neutron excess region) is defined as the difference between the rms radius of the neutron excess distribution, denoted as $R_{nexc}$, and the rms radius of proton, $R_p$, expressed as $\Delta R_{nexc} = R_{nexc} - R_p$. In $^{208}\text{Pb}$, core-polarization is considerable due to the large proton number. This leads to uncertainties in neutron distribution measurements due to the discrepancy between the $\rho_n - \rho_p$ and the $\rho_{nexc}$ distributions at the nuclear surface, which naturally arise from Coulomb interaction. In neutron distribution experiments, incident particles primarily interact with the region of neutron excess ($\Delta R_{nexc}$) rather than the neutron skin ($\Delta R_{np}$), potentially causing inaccuracies in determining the neutron skin thickness. Consequently, these uncertainties may be attributed to the core-polarization effect, underscoring the need to account for this phenomenon in theoretical analyses and the interpretation of experimental data.

The $(^{3}\text{He}, t)$ charge-exchange reaction at 420 MeV is commonly utilized to investigate the spin-isospin properties of nuclei \cite{Zegers07}. This reaction has effectively discovered neutron skin thickness, particularly through Fermi transitions between the isobaric analog state (IAS) and the target ground state.  Loc et al. \cite{Loc14} demonstrated this application by utilizing the IAS peak at zero scattering angle, and theoretical models provided an optimal Distorted Wave Born Approximation (DWBA) fit for the experimental data from Zegers et al. \cite{Zegers07}, enabling the determination of neutron skin thickness. This method was further corroborated in my previous work \cite{Huan21}, using the $G$-matrix double-folding model to determine nucleus-nucleus optical potential microscopically, emphasizing a non-negligible role of the isospin-density dependent component of $G$-matrix interaction in describing the experimental data. However, these studies did not account for the Coulomb core-polarization effect, an intrinsic feature that may influence theoretical analyses of this reaction.  

At this energy, 420 MeV, the $^{3}\text{He}$ particle predominantly probes the surface of the target and has been used to theoretically determine neutron skin thickness \cite{Loc14, Huan21}. Nevertheless, the intrinsic Coulomb core-polarization effect, which displaced the proton outward obviously, introduces a discrepancy between the neutron-proton difference and the neutron excess distributions in the surface region. As a result, at zero scattering angle, where the IAS peak is observed, the incident particle primarily probes the neutron excess region rather than the neutron skin. In nuclei with a small neutron excess, using the $\rho_{nexc}$ as the transition density instead of $\rho_n-\rho_p$ density to describe the IAS charge-exchange reaction results in a better experimental agreement, as the core polarization leads to a higher density encountered by the incident particle \cite{Loc17, Huan21, Huan23}. Recognizing the core polarization as an intrinsic property of nuclear structure is critical. If the core-polarization effect were diminished, the neutron excess distribution would be nearly identical to the difference between neutron and proton distribution. This illuminates the importance of accounting for the core-polarization behavior when determining neutron skin thickness theoretically using  $(^{3}\text{He}, t)$IAS reaction.
 
I described the $(^{3}\text{He}, t)$IAS reaction within the framework of Lane model \cite{Lane62} and DWBA, utilizing a double-folding model \cite{Khoa00,Khoa14,Toyokawa15}, which incorporates two critical ingredients: the Chiral three-nucleon forces (3NFs) $G$-matrix effective interaction obtained from Brueckner-Hartree-Fock calculations in nuclear matter \cite{Toyokawa18}, and nuclear densities calculated using the Skyrme Hartree-Fock (SHF) model, to determine the nucleus-nucleus optical potential. To investigate the Coulomb core-polarization impact in the description of this reaction, I employed two options of transition density: the conventional isovector density ($\rho_n-\rho_p$) and the neutron excess density ($\rho_{nexc}$). This approach provides a fully microscopic description of the $(^{3}\text{He}, t)$IAS reaction without adjusting any free parameters. For details of the calculations, see the Supplementary Material \cite{Suppl}. Additionally, the densities of the $^{3}\text{He}$ and triton particles are derived from a three-body calculation using the Argonne v18 interaction \cite{Nielsen01}. Target nuclear densities are obtained from SHF calculations \cite{Colo13} using the SAMi-J EDF family \cite{Roca13}. The SAMi interaction, recognized for its improved spin-isospin properties, is specifically fitted to the properties of $^{208}\text{Pb}$ \cite{Roca12}, ensuring a consistent mean-field description. As summarized in Table \ref{208Pb}, by varying the symmetry energy, I simulated different neutron skin values in $^{208}\text{Pb}$, covering a range consistent with the experimental neutron skin thickness from RCNP, PREX-II, and reanalyzed PREX-II results, providing a comprehensive basis for assessing the core-polarization process respect the increasing of neutron skin. The double-folding model, integrated with Chiral 3NFs $G$-matrix and SAMi-derived densities, offers a fully microscopic framework ideally compatible with this analysis.

\begin{table}[h!]
\caption{Properties of the $^{208}\text{Pb}$ obtained from Skyrme Hartree-Fock calculations using the SAMi-J interaction. Neutron skin thickness $\Delta R_{np} = R_n -R_p$, neutron excess thickness $\Delta R_{nexc} = R_{nexc} -R_p$, and $\Delta R_{core} = R_{ncore} - R_{pcore}$ that reflected the Coulomb core-polarization, are listed for different SAMi-J interactions. All rms radius are in fm.}
	\begin{tabular}{cccccccc}
		\hline
		\hline 
		SAMi  & $R_n$ & $R_{p(pcore)}$ & $\Delta R_{np}$  & $R_{ncore}$ & $\Delta R_{core}$ & $R_{nexc}$ & $\Delta R_{nexc}$ \\
		\hline 
		J29 & 5.625 & 5.463 & 0.162 & 5.321 & -0.142 &  6.150 & 0.687 \\
		J30 & 5.647 & 5.467 & 0.180 & 5.344 & -0.123 &  6.173 & 0.706 \\
		J31 & 5.668 & 5.469 & 0.199 & 5.366 & -0.103 &  6.193 & 0.724 \\
		J35 & 5.728 & 5.462 & 0.266 & 5.428 & -0.034 &  6.249 & 0.787 \\
		\hline		
		\hline
		\label{208Pb}
	\end{tabular}
\end{table}

\begin{figure}[h!]
	\centering
	\includegraphics[scale=0.44]{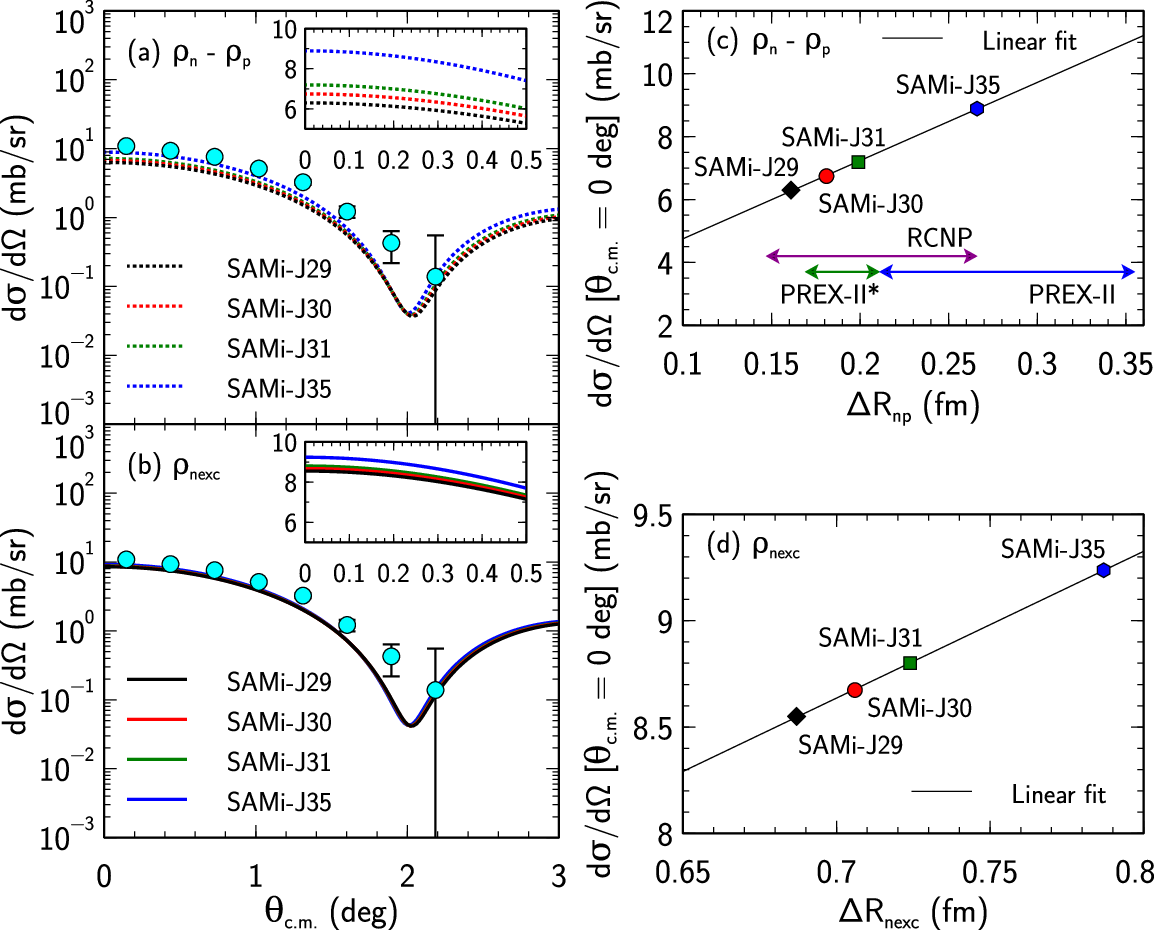}
	\caption{Differential cross section of the $^{208}\text{Pb}(^{3}\text{He}, t)$IAS reaction with SAMi-J interactions, using (a) isovector ($\rho_n-\rho_p$) and (b) neutron excess ($\rho_{nexc}$) transition densities. Two insets are provided to show the differential cross sections on a linear scale for each case, clearly illustrating the discrepancies at forward angles from 0$^\circ$ to 0.5$^\circ$. Correlation between the differential cross section at $0^{\circ}$ angle versus (c) neutron skin thickness using isovector density and (d) neutron excess thickness using neutron excess density. Experimental data are from Zegers et al. \cite{Zegers07}. These results emphasize the sensitivity of the $(^{3}\text{He}, t)$IAS reaction to neutron excess at forward angles due to the core-polarization influence. PREX-II* refers to the reanalyzed results of PREX-II \cite{Reinhard21}.}
	\label{0deg}
\end{figure}

To validate the core-polarization effect, I demonstrate that the $(^{3}\text{He}, t)$IAS reaction predominantly probes the neutron excess region, rather than the neutron skin, at forward scattering angles. This is particularly relevant at zero scattering angle, where the IAS peak is observed. My findings highlight that, at these angles, the $(^{3}\text{He}, t)$IAS reaction primarily probes neutron excess region. This is illustrated in Fig.~\ref{0deg}, which shows the differential cross section of the $^{208}\text{Pb}(^{3}\text{He}, t)$IAS reaction using two transition density options: the $\rho_n-\rho_p$ [Fig.~\ref{0deg}(a)] and the $\rho_{nexc}$ [Fig.~\ref{0deg}(b)], with the SAMi-J29 to SAMi-J35 interactions. Although all calculations qualitatively capture the general features of the cross sections, those employing $\rho_{nexc}$ [Fig.~\ref{0deg}(b)] provide the most accurate quantitative description of the experimental data. Additionally, I included two insets showing the differential cross sections on a linear scale for each case to clearly illustrate the discrepancies at forward angles from $0^\circ$ to $0.5^\circ$. I emphasize that using the same Skyrme interaction that yields a specific neutron skin value, the results derived from the $\rho_{nexc}$ density provide a more accurate description of the experimental data compared to those obtained from the $\rho_n-\rho_p$ density, due to the interaction of the $^3$He particle with the denser neutron region resulting from core polarization, demonstrates this reaction is primarily sensitive to the neutron excess region. In addition, as the neutron skin increases from SAMi-J29 to SAMi-J35, the differential cross section at forward angles is enhanced with both transition densities. Besides, Figure~\ref{0deg}(c) shows the correspondence between neutron skin thickness ($\Delta R_{np}$) and the differential cross section at exactly zero scattering angle using $\rho_n-\rho_p$ density, demonstrating a linear association. Conversely, Figure~\ref{0deg}(d) illustrates the correlation between the differential cross section at $0^{\circ}$ angle using $\rho_{nexc}$ density, and the neutron excess thickness ($\Delta R_{nexc}$), showing a similar linear trend that reflects the sensitivity of this reaction with the neutron excess distribution. Skyrme interactions that predict an increased $\Delta R_{np}$ also exhibit larger values of $\Delta R_{nexc}$ (see Table~\ref{208Pb}), which aligns with the observed increase in the differential cross section at a $0^{\circ}$ angle when the $^{3}\text{He}$ particle interacts with the greater $\Delta R_{nexc}$. All of the discussions mentioned above demonstrate that the $(^{3}\text{He}, t)$IAS reaction, which is sensitive to the neutron excess at forward angles due to the core-polarization phenomenon, probes the neutron excess region rather than the neutron skin at the surface, thereby underscoring the necessity of validating this effect.

\begin{figure*}[htb]
	\centering
	\includegraphics[scale=0.53]{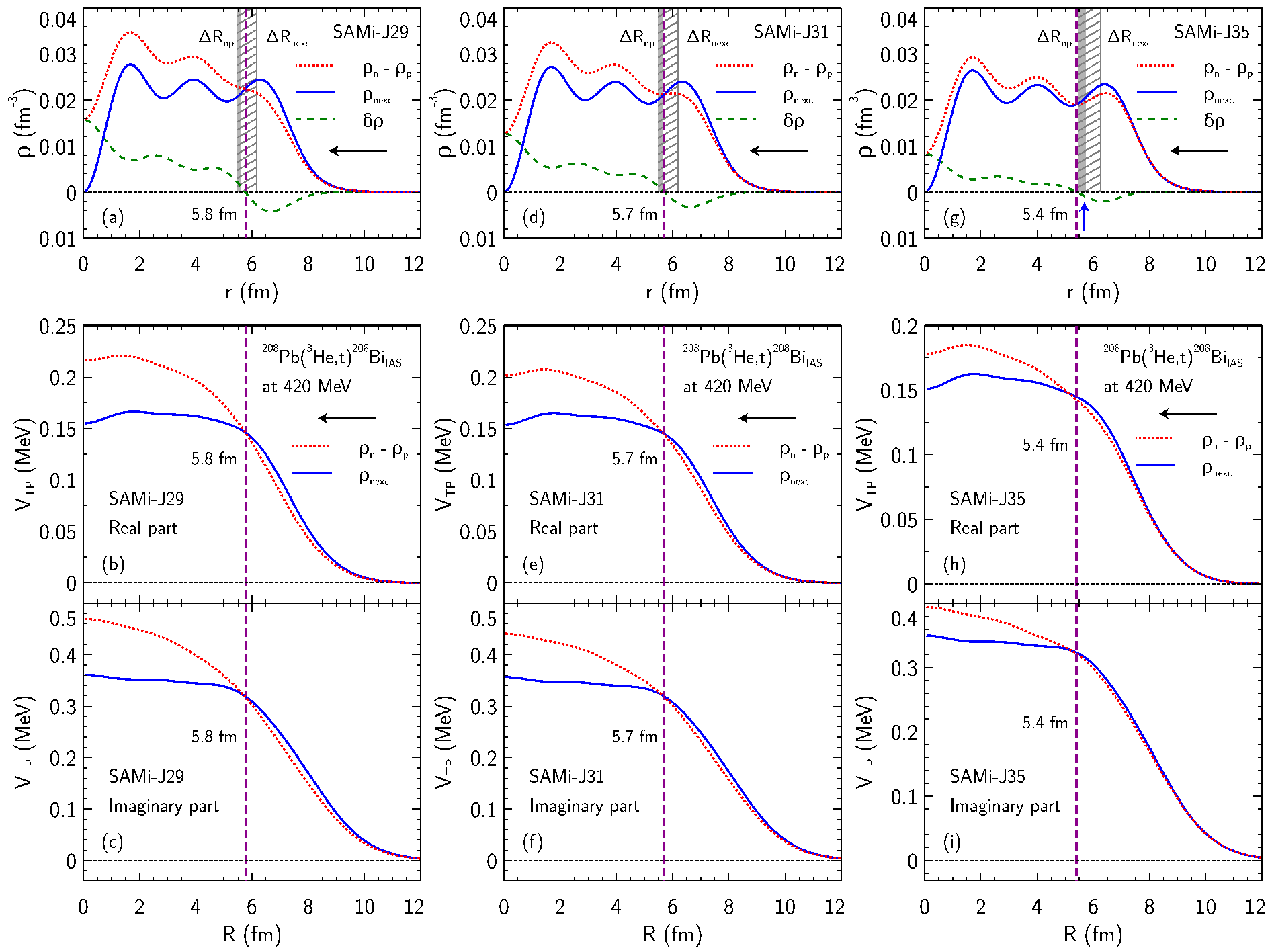}
	\caption{Nuclear distributions of $^{208}\text{Pb}$ for SAMi-J29, SAMi-J31, and SAMi-J35, illustrating isovector ($\rho_n - \rho_p$), neutron excess ($\rho_{nexc}$), and Coulomb core-polarization ($\delta \rho$) densities in (a, d, g). Real and imaginary parts of the transition potential using $\rho_n - \rho_p$ and $\rho_{nexc}$ transition densities are shown in (b, c) for SAMi-J29, (e, f) for SAMi-J31, and (h, i) for SAMi-J35. The divergence between $\rho_{nexc}$ and $\rho_n - \rho_p$ densities at the nuclear surface is featured. The neutron skin thickness ($\Delta R_{np}$) is displayed by the grey area, and neutron excess thickness ($\Delta R_{nexc}$) is exhibited by the forward-slashed area. The direction of $^{3}\text{He}$ particle probing the $^{208}\text{Pb}$ is marked by black horizontal arrows, and the Coulomb boundary radius is displayed by purple dashed lines. The blue vertical arrow in the case of SAMi-J35 (g) denotes where protons are mixed into the neutron skin area.}
	\label{TransPotDens}
\end{figure*}

\begin{figure}[h!]
	\centering
	\includegraphics[scale=0.5]{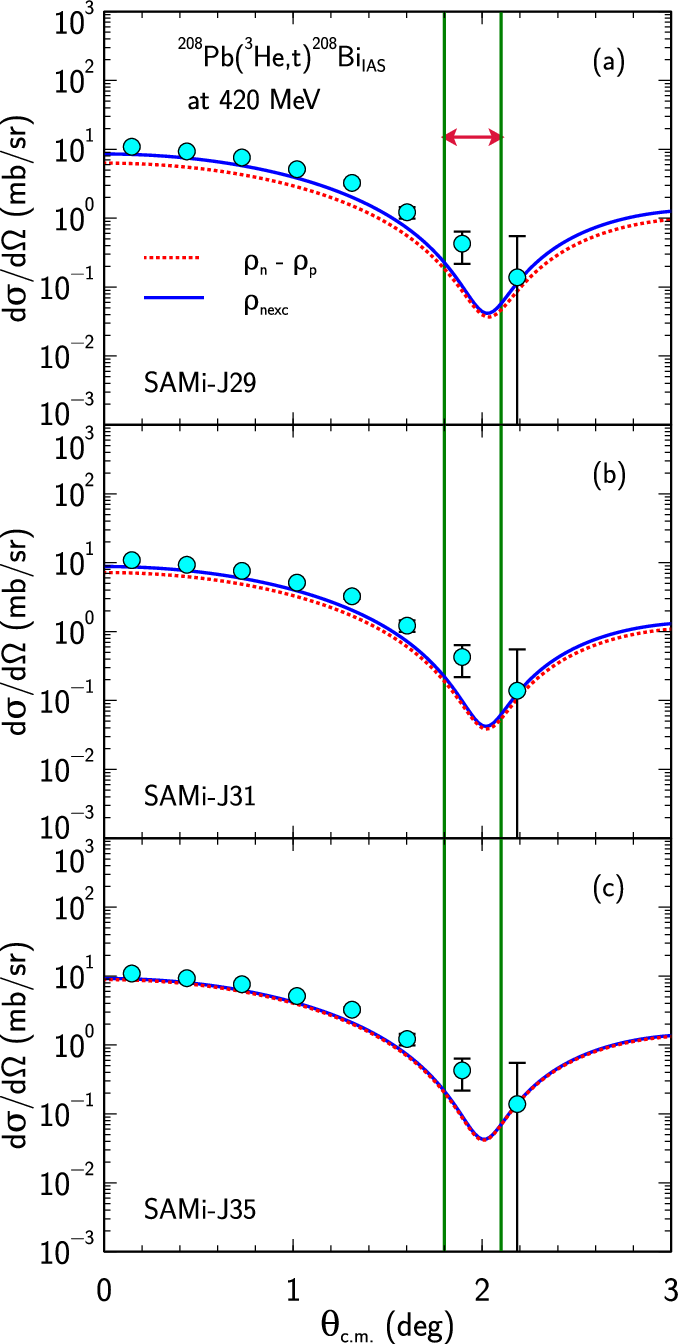}
	\caption{Differential cross section of the $^{208}\text{Pb}(^{3}\text{He}, t)$IAS reaction using isovector and neutron excess densities for different SAMi-J interactions: SAMi-J29 (a), SAMi-J31 (b), and SAMi-J35 (c). The red horizontal arrow indicates the region where the incident particle reaches the Coulomb boundary radius, with Coulomb core-polarization becoming negligible, approximately near the first minimum, between $1.8^\circ$ and $2.1^\circ$. Experimental data are from Zegers et al. \cite{Zegers07}.}
	\label{Min}
\end{figure}

I further elucidate the underlying physics from both nuclear structure and nuclear reaction perspectives. Figures \ref{TransPotDens}(a,d,g) display the nuclear distribution of $^{208}\text{Pb}$ for SAMi-J29, SAMi-J31, and SAMi-J35, respectively, illuminating the isovector ($\rho_n-\rho_p$), neutron excess ($\rho_{nexc}$), and Coulomb core-polarization ($\delta \rho$) densities. A clear deviation between the $\rho_{nexc}$ and $\rho_n-\rho_p$ density is observed at the nucleus surface, where protons are pushed toward, resulting in a negative density of $\delta \rho$ in the surface region. When the $^{3}\text{He}$ probes the target (exhibited by a black horizontal arrow), it first encounters the neutron excess region, characterized by the neutron excess thickness ($\Delta R_{nexc}$), displayed by the forward-slashed area. This interaction, as demonstrated in Fig. \ref{0deg}, occurs at forward angles. Figures \ref{TransPotDens}(b,c), \ref{TransPotDens}(e,f), and \ref{TransPotDens}(h,i) show the real and imaginary parts of the transition potential, which identify the differential cross section of the $^{208}\text{Pb}(^{3}\text{He}, t)$IAS reaction, using isovector and neutron excess transition densities. The transition potential characterizes the reaction probability and directly influences the differential cross section. The concept of the Coulomb boundary radius where protons begin to be displaced outward, is displayed by purple dashed lines. As the particle probes the target, it first interacts with the neutron excess area, resulting in an enhanced transition potential when using neutron excess density. This consistent behavior in both real and imaginary parts, underscores the microscopic nature of this model, with significant contributions from the imaginary part of optical potential coupling. The Coulomb boundary radius values are found to be 5.8 fm for SAMi-J29, 5.7 fm for SAMi-J31, and 5.4 fm for SAMi-J35. The same behavior in transition potential, including the point where the two transition potentials converge, is observed at a radius almost exactly equal to the Coulomb boundary radius. At the boundary radius, where the core-polarization impact vanishes, resulting in similar transition potentials for both $\rho_{nexc}$ and $\rho_n-\rho_p$ densities, consistent in both real and imaginary parts.

In fundamental physics, neutron skin is created by the symmetry potential, arising naturally from the asymmetry between protons and neutrons in the nucleus. However, the Coulomb core-polarization phenomenon, an intrinsic property, cannot be ignored. The similar strength of the transition potential at the convergence point denotes nearly the same reaction probability. Figure \ref{Min} presents the differential cross sections of the $^{208}\text{Pb}(^{3}\text{He}, t)$IAS reaction, using both the $\rho_n - \rho_p$ and $\rho_{nexc}$ transition densities for SAMi-J29 (a), SAMi-J31 (b), and SAMi-J35 (c). The red horizontal arrow in the figure indicates the region where the incident particle reaches the Coulomb boundary radius, where the influence of the core-polarization becomes negligible, reflected in the differential cross section. A specific overlap region between the differential cross sections for both transition potentials is identified at the first minimum, approximately between 1.8$^\circ$ and 2.1$^\circ$. Furthermore, by exploring the effect of core-polarization with increasing neutron skin values, simulated by different SAMi interactions, one would observe that the Coulomb boundary radius shifts from 5.8 fm to 5.4 fm as the neutron skin increases. The extent of protons driven outward due to core-polarization decreases with increasing neutron skin, as the symmetry potential becomes more dominant, as exhibited in Figs. \ref{TransPotDens}(a,d,g). This predominance is evident from the shifting Coulomb boundary radius and the reduction in $\Delta R_{core}$, which becomes less negative, as shown in Table \ref{208Pb}. The discrepancy between $R_{nexc}$ and $R_{n}$, caused by the influence of both the Coulomb and symmetry potential, is greater than $\Delta R_{core}$, which is affected solely by the Coulomb interaction. This shift and reduction in the number of protons driven outward highlight the diminished effect of core polarization as the neutron skin increases, as clearly observed in SAMi-J35, which exhibits the largest neutron skin value. This reduced effect results in minimal divergences in the transition potentials and corresponding differential cross sections obtained with the two transition densities.

Examining Figs. \ref{TransPotDens}(a,d,g), one would observe the density distribution with the neutron skin region $\Delta R_{np}$ displayed by the grey area. For SAMi-J29, with a neutron skin value of 0.162 fm, the $^{3}\text{He}$ particle first probes the neutron excess distribution, as indicated by the differential cross section at forward angles. Then, the particle reaches the Coulomb boundary radius, where core-polarization is unnoticeable, corresponding to the cross section at the first minimum (1.8$^\circ$ to 2.1$^\circ$). Here, the neutron skin region is positioned beyond the Coulomb boundary radius, meaning the particle naturally probes the neutron skin in $^{208}\text{Pb}$ beyond the first minimum. A similar trend is observed for SAMi-J31, with an increased neutron skin value of 0.199 fm. The Coulomb boundary radius shifts inward to 5.7 fm due to the symmetry potential. Thus, the scattering angle used to probe the neutron skin remains consistent with the SAMi-J29 case, occurring after the first minimum angle. In the case of SAMi-J35, with a neutron skin value of 0.266 fm, nearly matching the PREX-II result, an intriguing behavior is observed. The high neutron skin value indicates a predominant symmetry potential, shifting the Coulomb boundary radius further inward. Interestingly, the Coulomb boundary radius currently is located after the neutron skin region. This results in a mixed region where protons are driven into the neutron skin area (exhibited by a blue vertical arrow in Fig. \ref{TransPotDens}(g)), causing the impurity of neutron skin due to the core-polarization influence. In this scenario, after encountering the neutron excess, the $^{3}\text{He}$ particle probes an impure neutron skin region.

\begin{figure}[h!]
	\centering
	\includegraphics[scale=0.44]{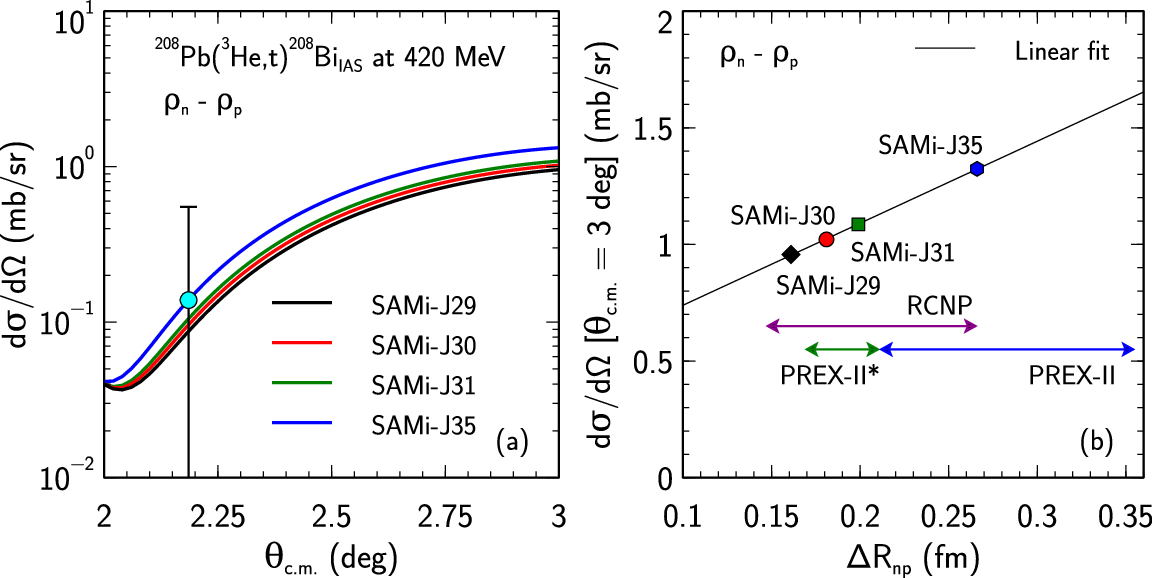}
	\caption{(a) Differential cross section of the $^{208}\text{Pb}(^{3}\text{He}, t)$IAS reaction at scattering angles from 2$^\circ$ to 3$^\circ$ using SAMi-J interactions incorporating isovector density. (b) Linear correlation between the differential cross section at exactly 3$^\circ$ and the neutron skin thickness for the same interactions. These results feature the sensitivity of the $(^{3}\text{He}, t)$IAS reaction to neutron skin at 3$^\circ$, where the core-polarization is negligible. Experimental data are from Zegers et al. \cite{Zegers07}.}
	\label{3deg}
\end{figure}

The $^{3}\text{He}$ behavior, as previously discussed, is consistent if the neutron skin of $^{208}\text{Pb}$ matches in case of the SAMi-J29 and SAMi-J31. My analysis currently centers on the scattering angle of 3$^\circ$. Figure \ref{3deg}(a) presents the differential cross section from 2$^\circ$ to 3$^\circ$ using the same interactions as in Fig. \ref{0deg} and employing the isovector transition density. In this angle region, the core-polarization are insignificant, allowing the isovector density used to determine the neutron skin thickness. As the neutron skin increases with different interactions, Fig. \ref{3deg}(b) illustrates the differential cross section at exactly 3$^\circ$ versus the neutron skin thickness. A linear trend similar to that in Fig. \ref{0deg}(c) is observed, demonstrating that the $(^{3}\text{He}, t)$IAS reaction is sensitive to the neutron skin thickness at 3$^\circ$, which is crucial for reliable determination.

It is vital to stress that the $(^{3}\text{He}, t)$IAS reaction probes the neutron excess at 0$^\circ$ angle due to the core-polarization process. However, once the incident particle traverses the Coulomb boundary, where this effect becomes negligible, it accurately probes the neutron skin. Therefore, the scattering angle of 3$^\circ$ is optimal for determining the precise neutron skin thickness in $^{208}\text{Pb}$. Theoretical models that reproduce the best quantitative description of the experimental data at scattering angles between 2$^\circ$ and 3$^\circ$ can correctly deduce the neutron skin thickness. I propose extending the $^{208}\text{Pb}(^{3}\text{He}, t)$IAS experiment by Zegers et al. \cite{Zegers07} at the Facility for Rare Isotope Beams (FRIB) to measure the differential cross section up to 3$^\circ$, where the exact neutron skin thickness can be resolved without the interference of core-polarization. This method is practical and allows for direct comparison with the PREX-II result. The advanced facilities at FRIB enable isobaric single charge-exchange reactions to be measured using exotic beams in inverse kinematics, allowing for measurements with minimal uncertainty.

The author is grateful to Prof. Dao Tien Khoa for valuable support and discussions. The author would also like to thank Dr. Bui Minh Loc for valuable suggestions. The author was funded by the PhD Scholarship Programme of Vingroup Innovation Foundation (VINIF), code VINIF.2024.TS.018.

\bibliographystyle{apsrev4-2-author-truncate}
\bibliography{Skin-Impurity}

\end{document}


\title{Supplemental Material for \\ ``Neutron skin impurity from Coulomb core polarization in $^{208}\text{Pb}$: \\ Insights from PREX-II and validation via the $(^{3}\text{He}, t)$ isobaric analog state reaction''}

\author{Phan Nhut Huan} 
\email{phannhuthuan@duytan.edu.vn}
\affiliation{Institute of Fundamental and Applied Sciences, Duy Tan University, Ho Chi Minh City 70000, Vietnam}
\affiliation{Faculty of Natural Sciences, Duy Tan University, Da Nang City 50000, Vietnam}

\begin{abstract}
This supplemental material provides a brief review of the formalism used to study the charge-exchange $(^{3}\text{He}, t)$ isobaric analog state (IAS) reaction within the $G$-matrix double-folding model, utilizing two options for the transition density: conventional isovector density, $\rho_n - \rho_p$, and neutron excess density, $\rho_{nexc}$, to explore the effect of Coulomb core polarization. \\
\end{abstract}

\maketitle

I summarize the formalism applied in my analysis of the $(^{3}\text{He}, t)$ isobaric analog state (IAS) reaction. Within the framework of the Lane model \cite{Lane62}, the central part of nucleus-nucleus optical potential is expressed as
\begin{equation}
	U(E,\bm{R}) = U_0(E,\bm{R}) + 4U_1(E,\bm{R}) \frac{\bm{T_a} \cdot \bm{T_A}}{aA},
\end{equation}
where $E$ is the total energy in the center-of-mass frame of the system, $\bm{R}$ is the relative coordinate vector, and $U_0(E,\bm{R})$ and $U_1(E,\bm{R})$ represent the isoscalar and isovector components of the optical potential, respectively. $\bm{T_a}$ and $\bm{T_A}$ are the isospin operators for the projectile and target, with $a$ and $A$ as their respective mass numbers.

In the Distorted Wave Born Approximation (DWBA), the transition amplitude for the charge-exchange isobaric analog state reaction is given by
\begin{equation}
	T^{\text{DWBA}}(E) = \langle \chi_{\tilde{a}\tilde{A}} | V_{\rm TP} (E,\bm{R}) | \chi_{aA} \rangle = -\frac{2\sqrt{2T_{z_A}}}{aA} \langle \chi_{\tilde{a}\tilde{A}} | U_1(E,\bm{R}) | \chi_{aA} \rangle,
\end{equation}
where $T_{z_A} = (N - Z)/2$ is the isospin's third component in the target's ground state. $ V_{\rm TP} (E,\bm{R}) = -\dfrac{2\sqrt{2T_{z_A}}}{aA} U_1(E,\bm{R})$ denotes the transition potential of this reaction. The distorted wave functions for the incoming ($\chi_{aA}$) and outgoing ($\chi_{\tilde{a}\tilde{A}}$) particles are obtained from the Schr\"{o}dinger equations
\begin{equation}
	\left[ K_a + U_0(E,\bm{R}) - \frac{2T_{z_A}}{aA} U_1(E,\bm{R}) + V_C(\bm{R}) - E_a \right]\chi_{aA}(\bm{R}) = 0
\end{equation}
and
\begin{equation}
	\left[ K_{\tilde{a}} + U_0(E,\bm{R}) + \frac{2(T_{z_A} - 1)}{aA} U_1(E,\bm{R}) + \Delta_C - E_a \right] \chi_{\tilde{a}\tilde{A}}(\bm{R}) = 0,
\end{equation}
where $K_{a(\tilde{a})}$ is the kinetic energy operator, $V_C(\bm{R})$ is the Coulomb potential, and $\Delta_C$ is the Coulomb displacement energy.

I used the double-folding model \cite{Toyokawa15, Khoa00} to determine the isoscalar and isovector parts of the nucleus-nucleus optical potential. The isoscalar $U_0(E,\bm{R})$ is determined as
\begin{equation}
	U_0(E,\bm{R}) = \iint \left[ \rho_a(\bm{r_a}) g_0^{\text{DR}}(E,\rho,s) \rho_A(\bm{r_A}) + \rho_a(\bm{r_a},\bm{r_a}+\bm{s}) g_0^{\text{EX}}(E,\rho,s) j_0\left(\frac{\bm{k}(E,\bm{R})\bm{s}}{M_{aA}}\right) \rho_A(\bm{r_A},\bm{r_A}-\bm{s}) \right] d\bm{r_a} d\bm{r_A}.
\end{equation}
The isovector $U_1(E,\bm{R})$, using the conventional isovector transition density $\rho_n - \rho_p$, is expressed as 
\begin{eqnarray}
	U_1(E,\bm{R}) &=& \frac{1}{\varepsilon} \iint \left[ \Delta \rho_a(\bm{r_a}) g_1^{\text{DR}}(E,\rho,s) \Delta \rho_A(\bm{r_A}) + \Delta \rho_a(\bm{r_a},\bm{r_a}+\bm{s}) g_1^{\text{EX}}(E,\rho,s) j_0\left(\frac{\bm{k}(E,\bm{R})\bm{s}}{M_{aA}}\right) \Delta \rho_A(\bm{r_A},\bm{r_A}-\bm{s}) \right] \nonumber\\ &\times& d\bm{r_a} d\bm{r_A},
	\label{TPIV}
\end{eqnarray}
where $\bm{s} = \bm{r_A} - \bm{r_a} + \bm{R}$, $\rho_i = \rho_i^n + \rho_i^p$ and $\Delta \rho_i = \rho_i^n - \rho_i^p$ are the isoscalar and isovector density matrices of the $i$-th nucleus, $\varepsilon = (N-Z)/A$ is the neutron-proton asymmetry of the target, $j_0$ is the zero-order spherical Bessel function, and $M_{aA} = aA/(a+A)$. The relative momentum $\bm{k}(E,\bm{R})$ is determined from
\begin{equation}
	k^2(E,\bm{R}) = \frac{2\mu}{\hbar^2} \left[ E - V(E,\bm{R}) - V_C(\bm{R}) \right],
\end{equation}
where $\mu$ is the reduced mass of the projectile-target system and $V(E,\bm{R})$ is the real part of the optical potential.

The isoscalar and isovector direct ($g_0^{\text{DR}}, g_1^{\text{DR}}$) and exchange ($g_0^{\text{EX}}, g_1^{\text{EX}}$) terms of the effective nucleon-nucleon interaction are defined by linear combinations of spin-isospin (ST) channels
\begin{eqnarray}
	g_0^{\text{DR}} &=& \frac{1}{16} \left[ g^{(00)} + 3 g^{(10)} + 3 g^{(01)} + 9g^{(11)} \right], \\
	g_1^{\text{DR}} &=& \frac{1}{16} \left[ -g^{(00)} - 3 g^{(10)} + g^{(01)} + 3g^{(11)} \right], \\
	g_0^{\text{EX}} &=& \frac{1}{16} \left[ -g^{(00)} + 3 g^{(10)} + 3 g^{(01)} - 9g^{(11)} \right], \\
	g_1^{\text{EX}} &=& \frac{1}{16} \left[ g^{(00)} - 3 g^{(10)} + g^{(01)} - 3g^{(11)} \right].
\end{eqnarray}
The form of $g^{(ST)}(E,\rho,s)$ in coordinate space is modeled by the three ranges Gaussian form
\begin{equation}
	g^{(ST)}(E,\rho,s) =\sum_{i = 1}^3 G^{(ST)}_{i}(E,\rho) \exp\left[-\left(\frac{s}{\lambda_i}\right)^2\right],
\end{equation}
where $G^{(ST)}_i(E,\rho)$ are complex energy-density dependent strengths, including both the real and imaginary parts, and $\lambda_i$ are the range parameters of the effective interaction. This study employs the three-nucleon forces (3NFs) version of the Chiral $G$-matrix interaction, with the specific parameter sets detailed in Ref. \cite{Toyokawa18}.

To investigate the core-polarization impact, the transition potential $V_{\rm TP}$ was calculated using two options for the transition density: the conventional isovector density, $\rho_n - \rho_p$, and the neutron excess density, $\rho_{nexc}$. The transition potential for $\rho_n - \rho_p$ is expressed in Eq. \eqref{TPIV}. For the $\rho_{nexc}$ transition density, I replaced the isovector density matrix of the target nucleus in Eq. \eqref{TPIV} with the corresponding $\rho_{nexc}$, while retaining $\rho_n - \rho_p$ for the $^3$He projectile, as expressed in
\begin{eqnarray}
	U^{nexc}_1(E,\bm{R}) &=& \frac{1}{\varepsilon} \iint \left[ \Delta \rho_a(\bm{r_a}) g_1^{\text{DR}}(E,\rho,s) \rho_{nexc}(\bm{r_A}) + \Delta \rho_a(\bm{r_a},\bm{r_a}+\bm{s}) g_1^{\text{EX}}(E,\rho,s) j_0\left(\frac{\bm{k}(E,\bm{R})\bm{s}}{M_{aA}}\right) \rho_{nexc} (\bm{r_A},\bm{r_A}-\bm{s}) \right] \nonumber \\ &\times& d\bm{r_a} d\bm{r_A}.
	\label{TPNEXC}
\end{eqnarray}
Finally, all one-body density matrices were localized following Ref. \cite{Brieva77II}. The nuclear densities for the targets are derived from Skyrme-Hartree-Fock calculations \cite{Colo13} using different SAMi-J interactions \cite{Roca12,Roca13}, while $^3$He and triton densities are obtained from three-body calculations using the Argonne v18 nucleon-nucleon interaction \cite{Nielsen01}. DWBA calculations were performed using the ECIS06 code \cite{Raynal06}. 

\bibliographystyle{apsrev4-2}
\bibliography{SM}